\documentclass[useAMS,usenatbib,usegraphicx]{mn2e}

\title[Is there a standard measuring rod in the Universe?]{Is there a standard
measuring rod in the Universe?}
\author[J. C. Jackson]{J. C. Jackson$^{1}$\thanks{E-mail:
john.jackson@unn.ac.uk}\\
$^{1}$Division of Mathematics, School of Computing, Engineering and Information Sciences,
Northumbria University, Newcastle NE1 8ST, UK\\
}
\begin{document}

\date{Accepted 2008 December 15. Received 2008 December 14; in original form 2008 October 11}

\pagerange{\pageref{firstpage}--\pageref{lastpage}} \pubyear{2008}

\maketitle

\label{firstpage}

\begin{abstract}
The Caltech–Jodrell Bank very long baseline interferometry (VLBI) Surveys give detailed
5 GHz VLBI images of several hundred milliarcsecond (mas) radio sources, and the full width
at half-maximum angular sizes of the corresponding compact cores. Using the latter, I have
constructed an angular-diameter/redshift diagram comprising 271 objects, which shows clearly
the expected features of such a diagram, without redshift binning. Cosmological parameters
are derived which are compatible with existing consensus values, particularly when the VLBI
data are combined with recent Baryon Acoustic Oscillations observations; the figures are
presented as indications of what might be expected of larger samples of similar data. The
importance of beaming and relativistic motion towards the observer is stressed; a model of the
latter indicates that the emitting material is close to the observer's line of sight and moving
with a velocity which brings it close to the observer's rest frame. With respect to linear size,
these objects compare reasonably well in variance with the absolute luminosity of type Ia
supernovae; the efficacy of the latter is improved by the brighter-slower and brighter-bluer
correlations, and by the inverse-square law.    
\end{abstract}


\section{Introduction}

Cosmological parameters are now known to a remarkable degree of precision, particularly 
those derived from {\it Wilkinson Microwave Anisotropy Probe (WMAP)} observations \citep{b9,b22}, 
in combination with observations of Type Ia supernovae (SNe Ia) \citep{b34,b1,b35,b42},
and the imprint of Baryon Accoustic Oscillations (BAO) on the distribution
of galaxies \citep{b4,b37,b10}; current values are $\Omega_{\rmn{m}}=0.279\pm 0.013$
and $\Omega_\Lambda=0.721\pm 0.015$ for the matter and vacuum parameters respectively (unless
otherwise noted, all confidence limits quoted here are 68 per cent ones).  Whereas the
{\it WMAP} and BAO approaches are of relatively recent origin, the magnitude/redshift approach has
a long and somewhat varied history, for want of a class of objects with similar absolue
magnitudes (see e.g. \citet{b43} for a review of early work); however, over the
last decade the latter approach has achieved spectacular success, with the discovery of
SNe Ia as accurate standard candles \citep{b33,b36,b27}.

In contrast, the angular-size/redshift approach has had {a surprisingly modest impact},
despite significant efforts in this direction, using extra-galactic radio sources in their
several guises as putative standard measuring rods.  Early work considered classical double
radio sources as suitable objects \citep{b23,b25,b20}, which approach continues to this day 
\citep{b7,b5,b8}.  Here I will re-examine ultra-compact radio sources as standard measuring
rods, with angular diameters in the milliarcsecond (mas) range, and linear sizes of order several
parsecs.  Their advantages in this context were first highlighted by \citet{b21}; these 
objects are much smaller than their parent active galactic nuclei, so that their
local environments should be similar and reasonably stable, at least over an appropriate redshift
range. They are energetic and shortlived, with central engines which are reasonably standard
objects (black holes with masses close to $1.5\times 10^{10}\rmn{M_\odot}$).  In these respects
they have much in common with SNe Ia, albeit with lifetimes of centuries rather than months. 
\citet{b21} presented angular sizes for a sample of 79 milliarcsecond (mas) sources,
obtained using very long baseline interferometry (VLBI) at 5GHz.  VLBI images tyically
show a compact core surrounded by debris, and \citet{b21} defined a characteristic
angular size as the distance between the core and the most distant component having
a peak brightness greater than or equal to 2 per cent of that of the core.
Typical linear sizes are $20 h^{-1}$ pc,\footnote{$H_0=100h$ km sec$^{-1}$ Mpc$^{-1}$}
and Kellermann noted that the corresponding angular-size/redshift diagram is compatible
with the once-favoured flat cold dark matter (CDM) model, $\Omega_{\rmn{m}}=1$,
$\Omega_\Lambda=0$. However, for a critique see Pearson et al. (1994); these authors
present a similar sample (with very little overlap) and use the same measure of angular
size, which shows virtually no change over the same redshift range.  \citet{b12} have
examined a much larger 5 GHz sample (330 objects), with results which do not resolve
this conflict. I suspect that the discord here is due to the definition of angular size,
which is too sensitive to the details of source structure.

\citet{b11} presented a large VLBI compilation, based upon a 2.29 GHz survey
undertaken by \citet{b32}; the compilation lists a rough measure of angular size 
based upon fringe visibility; this measure should be less sensitive to the details of 
source structure, and more representative of the compact core, which is usually the
dominant component with respect to  radio luminosity (see e.g. \citealt{b14}).
Gurvits considered a subset comprising 258 sources divided into 12 redshift bins over the
range $0.501\leq z\leq 3.787$, and found marginal support for a low-density CDM cosmological model,
assuming that selection and evolutionary effects can be ignored.  Gurvits considered only models
with $\Omega_\Lambda=0$. Using exactly the same data set (kindly supplied by Dr. Gurvits),
\citet{b18} extended the analysis to the full $\Omega_{\rmn{m}}$--$\;\Omega_\Lambda$ plane; 
the situation is very degenerate with respect to choice of $\Omega_\Lambda$, with both 
95 per cent and 68 per cent confidence regions extending well into the region $\Omega_\Lambda<0$.
Nevertheless, marginalizing over $\Omega_\Lambda>0$ gives $0.11\leq \Omega_{\rmn{m}}\leq 0.54$,
providing clear support for a low-density CDM model.  Imposing spatial flatness allowed
a much more definitive statement to be made: $0.1\la \Omega_{\rmn{m}}\la 0.3$, later refined
to $\Omega_{\rmn{m}}=0.24+0.09/-0.07$ (95 per cent confidence limits) \citep{b17}.
\footnote{Tighter figures are presented in \citet{b19}, where a larger sample is examined
(613 objects), produced by updating \citet{b32} with respect to redshift, and finding some
of the missing flux densities from elsewhere; this work places significant constraints
on $\Omega_{\rmn{m}}$ and $\Omega_\Lambda$ without assuming flatness.  However, in 
retrospect we have doubts about the suitabilty of the extra data, and about the efficacy 
of redshift binning, see discussion later; the precision in \citet{b19} has been overstated.}

\section{Data and Results}

The object here is to consider a more precise definition of angular size.
The Caltech-Jodrell Bank flat-spectrum (CJF) sample is a complete 5 GHz
flux-density-limited sample of 293 flat-spectrum sources, complete according
to the following criteria \citep{b39}: 

\begin{enumerate}
  \item Flux density at 4850 MHz $S_{4850}\geq 350$ mJy.\hfil\break
  \item Spectral index $\alpha\geq -0.5$ ($S\propto \hbox{frequency}^{\alpha}$).\hfil\break
  \item Declination $\delta\geq 35^\circ$ (1950 coordinates).\hfil\break
  \item Galactic latitude $\geq 10^\circ$.
\end{enumerate}

\begin{figure}
\vspace{85pt}
\includegraphics[width=84mm]{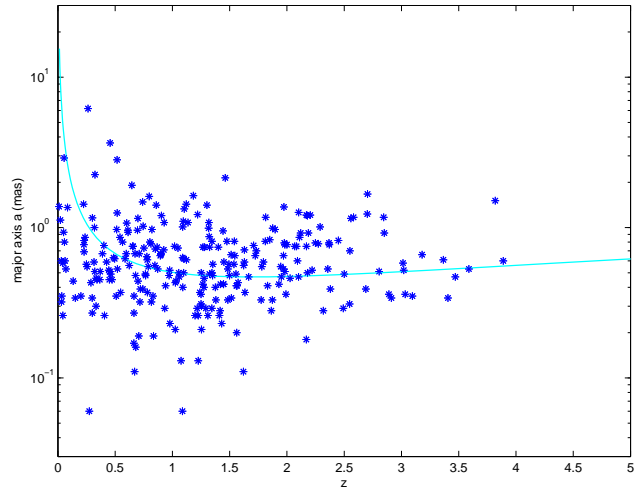}
\caption{Angular-diameter/redshift diagram for 271 sources from the composite
PR+CJ1+CJ2 sample.  {The cyan line corresponds to $\Omega_{\rmn{m}}=0.236$,
$\Omega_\Lambda=1-\Omega_{\rmn{m}}=0.764$ and $d=2.94 h^{-1}$ pc, see text}.}
\label{FigA}
\end{figure}

\citet{b39} present a VLBI image for each successfully imaged source, and give accurate full 
width at half-maximum angular major and minor axes ($a$ and $b$) for each component therein,
with the compact core clearly identified; the major axis of the latter will be taken
as the measure of angular size in this investigation.  In fact I have used a somewhat 
larger sample than the CJF one.  The latter is, in part, a subset of three earlier 
samples, the PR \citep{b28,b29}, CJ1 \citep{b31,b40,b44} and CJ2 \citep{b38,b14}
samples.  A composite comprising the latter three is nominally complete with respect
to criteria (i), (iii) and (iv) above, but not the spectral index one.  \citet{b39}
selected the CJF sample from this composite by imposing the spectral index limit,
and also added 18 further sources which had been missed in the earlier surveys,
which additional sources meet all of the above criteria.  In the interest of a modest
increase in numbers I have used the full composite sample PR+CJ1+CJ2, as listed in
tables 3 of \citet{b44} and \citet{b14} and tables 4 of \citet{b29} and \citet{b38},
plus 17 of the the 18 further sources, as listed in table 3 of \citet{b39}, giving
322 objects in all. Of these 6 are assigned major axes formally equal to 0.00
milliarcsec, presumably because they are  below the resolution limit of the VLBI
system; the said sources have been discarded.  I have updated the redshift 
list using the NASA/IPAC Extragalactic Database (NED), and for the remaining
316 objects find 271 redshifts. Fig. \ref{FigA} is a plot of major axis $a$ against
redshift $z$ for these sources.  Despite the spread the expected qualitative features
are {reasonably} clear, namely a diminishing size from $z=0$ to $z\sim 1$, folowed by a
gradual increase, with a minimum somewhere between $z\sim 1$ and $z\sim 1.5$, first
predicted by \citet{b15}.  The plot in Fig. \ref{FigA} does not appear to be resolution
limited.

\begin{figure}
\vspace{-95pt}
\end{figure}

\begin{figure}
\includegraphics[width=84mm]{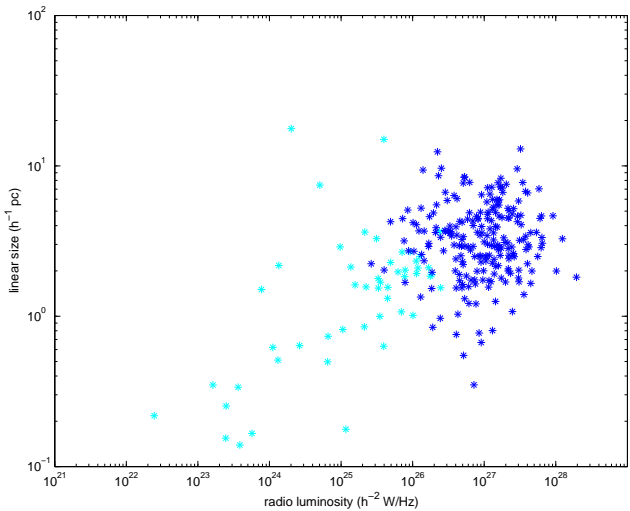}
\caption{Plot of linear size against radio luminosity for 271 sources from the
composite PR+CJ1+CJ2 sample; the cyan points have $z<0.5$, the blue ones $z\geq 0.5$.}
\label{FigB}
\end{figure}

Before turning to quantitative matters, a second diagram is instructive.
Fig. \ref{FigB} shows linear size plotted against radio luminosity for the 271
sources in Fig. \ref{FigA}, taking $\Omega_{\rmn{m}}=0.27$, $\Omega_\Lambda=0.73$.
The cyan points correspond to $z<0.5$, the blue ones to $z\geq 0.5$.
In a flux-limited sample sources observed at large redshifts are intrinsically 
the most powerful, so that a correlation between linear size and radio
luminosity would introduce a selection effect.  The high-redshift population
shows no obvious evidence of such a correlation, and gives every indication
of being statistically stable with respect to linear size.  In the low-redshift case 
this is clearly not so. Fig. \ref{FigB} gives a clear illustration of why
ultra-compact radio sources with $z\la 0.5$ are of no value in this context,
first noted by \citet{b11}, see also \citet{b18} and \citet{b17}.

In what follows I will work with data points which correspond to individual sources,
rather than putting the latter into redshift bins.  Although binning is popular and
can reveal trends in data which are not otherwise obvious to the eye, with regard to
quantitative statistical analysis there is, in principle, no advantage in such a procedure;
what is gained by having smaller error bars is lost by having fewer points.  Appearances
can be deceptive, particularly if numbers are small, and results can be overly sensitive
to the choice of bins.  I will concentrate on spatially flat $\Lambda$CDM models,
characterized by $\Omega_{\rmn{m}}$ and a characteristic linear size $d$ associated 
with the source population.  I will use objects in the range $0.5\leq z\leq 3.5$;
the upper limit on $z$ removes 3 points, one of which is an outlier which has an
inordinate effect upon the statistical analysis.

Giving each point equal weight and taking $\log a$ as ordinate, the best-fitting 
flat model is $\Omega_{\rmn{m}}=0.931$, $\Omega_\Lambda=0.069$ and $d=2.33 h^{-1}$ pc, close
to the erstwhile canonical model $\Omega_{\rmn{m}}=1$, $\Omega_\Lambda=0$.  However, there is
another parameter which I believe is an important source discriminator, namely the
axial ratio $r=b/a$.  This belief is based upon an astrophysical model, discussed at
length in Jackson (2004).  According to this model the underlying source population
consists of compact symmetric objects \citep{b41}, comprising central low-luminosity 
cores straddled by two mini-lobes.  The compact components which are the basis 
of this study are identified as cases in which the lobes are moving relativistically,
and are close to the line of sight, when D\H oppler boosting allows just that material
which is moving towards the observer to be seen.  As $z$ increases a larger
D\H oppler factor is required; it turns out that the latter approximately cancels the
cosmological redshift, so that the observed component is seen in its rest frame.
This is a very important effect, because the measured size of mas source components
is known to increase linearly with wavelength \citep{b24,b28}.  Without the said
effect mas angular-size/redshift diagrams would show something like the so-called
Euclidean behaviour, angular size proportional to $1/z$ (because the emitted frequency
is $(1+z)~\times$ the received one), which bedevilled early work on classical 
double radio sources \citep{b23,b25,b20,b16}.  It is quite remarkable that 
Fig. \ref{FigA} shows no trace of such behaviour, which fact is a striking
confirmation of the proposed model.

\begin{figure}
\includegraphics[width=84mm]{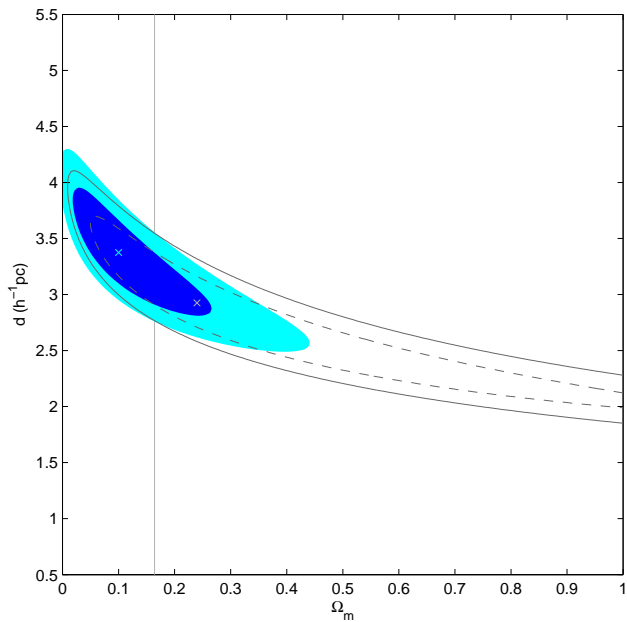}
\caption{Joint mas/BAO confidence regions in the $\Omega_{\rmn{m}}$--$\;d$ plane, 68 per cent (blue)
and 95 per cent (cyan); the mid-grey lines refer to mas sources alone; the light-grey vertical
line is the right-most 95 per cent BAO confidence limit.  Data relate to 128 sources from the composite PR+CJ1+CJ2 sample, with $z\geq 0.5$ and axial ratio $\geq 0.4$.}
\label{FigC}
\end{figure}

The ideal image will correspond to a head-on approach, and an axial ratio close to unity.  
I have examined samples which include only those objects with $r\geq r_c$,
and find that the best-fitting value of $\Omega_{\rmn{m}}$ depends upon the cut-off ratio
$r_c$ in a systematic fashion.  The said value falls from $0.931$ at $r_c=0$ to $0.253$ 
at $r_c=0.35$, and thereafter remains reasonably constant until we run out of sensible 
numbers: $\Omega_{\rmn{m}}=0.28\pm 0.04$ for $0.35\leq r_c\leq 0.6$, and $\Omega_{\rmn{m}}=0.29\pm 0.08$
for $0.35\leq r_c\leq 0.7$.  As a representative example I present results for the case 
$r_c=0.4$, $0.5\leq z\leq 3.5$, comprising 128 objects, for which the best figures 
are $\Omega_{\rmn{m}}=0.236$ and $d=2.94 h^{-1}$ pc; {the corresponding curve is shown in
Fig. \ref{FigA}}.  A fixed standard deviation $\sigma$ is attached to each point, being
defined by $\sigma^2=\hbox{residual sum-of-squares/}(n-p)$, where $n=128$ is the number of
points and $p=2$ is the number of fitted parameters; the appropriate value in this case is
\begin{equation}\label{A}
\sigma=0.252,
\end{equation} 
which value is used to calculate $\chi^2$ values at points in parameter space.
The mid-grey lines in Fig. \ref {FigC} show confidence regions in the $\Omega_{\rmn{m}}$--$\;d$
plane.  Marginalizing over $d$ gives $\Omega_{\rmn{m}}=0.24+0.40/-0.15$.

This somewhat indeterminate result is in large part occasioned by the absence of
standard objects in the redshift range $0<z<0.5$, which would otherwise determine
a model-independent normalization of the characteristic linear size $d$.  The
deficiency can be remedied by combining recent BAO observations with the above data.
I have in mind measures of the BAO scales at $z=0.2$ and $z=0.35$ \citep{b26},
which constrain values of the hybrid distance $D_V(z)=[(1+z)^2 D_A^2(z)cz/H(z)]^{1/3}~$, 
giving $D_V(0.35)/D_V(0.2)=1.812\pm 0.060$, which ratio does not refer directly to the
size of the acoustic horizon at recombination.  The joint mas+BAO confidence regions
are represented by the filled blue/cyan areas in Fig. \ref {FigC}.  Marginalizing
over $d$ gives $\Omega_{\rmn{m}}=0.10+0.09/-0.06$.  The light-grey vertical line is the
right-most 95 per cent BAO confidence limit; the rest are formally at negative
values of $\Omega_{\rmn{m}}$, including the best-fitting line $\Omega_{\rmn{m}}=-0.13$.

\section{Coda and Conclusions}

Ultra-compact radio sources do comprise a reasonably standard unit of linear size,
and in conjunction with BAO they are close to being useful for cosmological
investigations.  How well do they compare with SNe Ia in this respect?
Given a radio source at redshift $z$ with linear size $d$ and angular size
$\theta$, its luminosity distance is

\begin{equation}\label{B}
d_L=\left({d \over \theta}\right)(1+z)^2.
\end{equation}

\noindent
As apparent and absolute magnitudes are related by
\begin{equation}\label{C}
m=M+5(\log_{10}d_L-1)~~~(d_L\hbox{ in parsecs}),
\end{equation}

\noindent
an equivalent mock supernova is defined by

\begin{equation}\label{D}
m_{\rmn{mas}}=M+5[\log_{10}d+2\log_{10}(1+z)-\log_{10}\theta-1].
\end{equation}

\noindent
{A mock magnitude/redshift diagram is produced by assigning
approprate fixed values to $M$ and $d$, as in \citet{b19}}.  Thus 
a variation $\delta\theta$ at fixed $z$ is equivalent to a variation

\begin{equation}\label{E}
\delta m_{\rmn{mas}}=-5\,\delta\log_{10}\theta.
\end{equation}

\noindent
The equivalent dispersion is thus $\delta m_{\rmn{mas}}\sim \pm 1.26$
mag, from equation (\ref{A}).  For SNe Ia the observed figure is
$\delta m_{\rmn{SN}}\sim \pm 0.38$ mag, believed to be largely intrinsic
rather than photometric \citep{b45,b46}. The present mas performance 
thus appears to be inferior by a factor of about $3$.  In fact SNe Ia do
rather better than this, because their effective dispersion is reduced
by the brighter-slower and brighter-bluer correlations \citep{b47,b48,b13},
typically to $\delta m_{\rmn{SN}}\sim \pm 0.20$ mag \citep{b1} and
$\delta m_{\rmn{SN}}\sim \pm 0.27$ mag (\citealt{b34}, the gold subset).
The latter figures are believed to comprise photometric and residual
intrinsic errors in roughly equal proportions.  It is instructive to
look at the basic physical parameters in the two cases: the linear
size $d$ and absolute luminosity $L$.  The variation in $L$
is given by equation (\ref{C}) as

\begin{equation}\label{F}
{\delta L \over L}= -\log_e 10\times 0.4\times \delta m_{\rmn{SN}},
\end{equation}

\noindent
the equivalent for mas sources being

\begin{equation}\label{G}
{\delta d \over d}=\log_e 10\times \delta(\log_{10}\theta)=-\log_e 10\times 0.2\times \delta m_{\rmn{mas}}.
\end{equation}

\noindent
At first sight equations (\ref{F}) and (\ref{G}) present a puzzle; why, for equivalent
intrinsic variations $\delta L/L$ and $\delta d/d$, is $\delta m_{\rmn{mas}}$
larger than $\delta m_{\rmn{SN}}$ by a factor of $2$?  The answer is that it is the
inverse-square law, implicit in equations (\ref{C}) and (\ref{D}), which puts the angular
approach at a disadvantage.  The numerical values are not dissimilar: $\delta d/d\sim 0.58$
and $\delta L/L\sim 0.35$; the latter is reduced by a factor of about $1.6$ by the
above-mentioned correlations, whereas the former is in effect increased by a factor of
$2$ by the inverse-square law.
    
I have looked at spectral index $\alpha$ as a possible size discriminator for
mas sources, using the integrated values listed in \citet{b39}, but find that
the correlation is too weak to be useful, {over the range of $\alpha$ values
encompassed here.  The central black hole mass $M_{\rmn{bh}}$ would be an interesting
parameter in this context, which could be determined by the correlation between
$M_{\rmn{bh}}$ and velocity dispersion of the the host elliptical galaxy \citep{b50,b51},
or that between $M_{\rmn{bh}}$ and the S\'ersic luminosity concentration index of the
galaxy \citep{b52,b49,b53}.  The appropriate information is not currently available;
its acquisition would be a worthwhile undertaking}.  The variance in measured
size probably has significant contributions from both intrinsic and instrumental
effects, so that an increase in relolving power would improve matters.  There is
certainly scope for increasing the sample size; the Caltech-Jodrell Bank surveys
cover only about 20 per cent of the full sky, so that a five-fold increase would
be possible without changing the flux limit.  The comoving volume encompassed
by a complete survey of flat-spectrum sources with flux-density limit $S_l$
increases roughly as $S_l^{-1}$, over the redshift range considered here.

The mas/BAO combination is a natural one, in that it allows
cosmological parameters to be determined by data which are local ($z\la 4$)
and exclusively angular.  Any discernible differences between parameters so
determined and those determined by supernovae would be a manifestation of
differential selection or evolutionary effects, or of effects relating to
the astrophysics or possibly the fundamental physics of light propagation
over cosmological distances \citep{b2,b3,b6}.  On more general grounds a new 
approach is always of some value, even if its weight is relatively low, because
its systematic errors become random ones when the new technique is added to
an ensemble of existing ones.

Interested parties can obtain copies of the data set used in this investigation
by sending an email request to john.jackson@unn.ac.uk.

\section*{Acknowledgments}

This research has made use of the NASA/IPAC Extragalactic Database (NED) which is operated
by the Jet Propulsion Laboratory, California Institute of Technology, under contract with the
National Aeronautics and Space Administration.


\label{lastpage}


\begin{thebibliography}{}
\bibitem[\protect\citeauthoryear{Astier et al.}{2006}]{b1} Astier et al., 2006, A\&A, 447, 31
\bibitem[\protect\citeauthoryear{Bassett \& Kunz}{2004a}]{b2} Bassett B.A., Kunz M., 2004a,
      Phys. Rev., D69, 101305
\bibitem[\protect\citeauthoryear{Bassett \& Kunz}{2004b}]{b3} Bassett B.A., Kunz M., 2004b,
      ApJ, 607, 661
\bibitem[\protect\citeauthoryear{Blake \& Glazebrook}{2003}]{b4} Blake C., Glazebrook K., 2003,
      ApJ, 594, 665
\bibitem[\protect\citeauthoryear{Buchalter et al.}{1998}]{b5} Buchalter A., Helfand D.J.,
      Becker R.H., White R.L., 1998, ApJ, 494, 503
\bibitem[\protect\citeauthoryear{Burrage}{2008}]{b6} Burrage C., 2008, Phys. Rev., D77, 043009 
\bibitem[\protect\citeauthoryear{Daly}{1994}]{b7} Daly R.A., 1994, ApJ, 426, 38
\bibitem[\protect\citeauthoryear{Daly et al.}{2007}]{b8} Daly R.A., Mory M.P., O'Dea C.P.,
      Kharb P., Baum S., Guerra E.J., Djorgovski S.G., 2007, ApJ, in press (arXive 0710.5112)
\bibitem[\protect\citeauthoryear{Driver et al.}{2006}]{b49} Driver S.P., Graham A.W.,
      Allen P.D., Liske J., 2006, in LeBrun V., Mazure A., Arnouts S., Burgarella D., eds,
      Proc. Vth Marseille International Cosmology Conference, The Fabulous Destiny of Galaxies:
      Bridging Past and Present.  Frontier Group, Paris, p. 409
\bibitem[\protect\citeauthoryear{Dunkley et al.}{2008}] {b9} Dunkley J. et al., 2008, ApJS, in press
      (arXive 0803.0586)
\bibitem[\protect\citeauthoryear{Eisenstein et al.}{2005}]{b10} Eisenstein D.J. et al., 2005,
      ApJ, 633, 560
\bibitem[\protect\citeauthoryear{Ferrarese \& Merritt}{2000}]{b50} Ferrarese L., Merritt D., 2000, 
      ApJ, 539, L9
\bibitem[\protect\citeauthoryear{Gebhardt et al.}{2000}]{b51} Gebhardt K. et al., 2000, 
      ApJ, 539, L13
\bibitem[\protect\citeauthoryear{Graham \& Driver}{2007}]{b53} Graham A.W., Driver S.P.,
      2007, ApJ, 655, 77
\bibitem[\protect\citeauthoryear{Graham et al.}{2001}]{b52} Graham A.W., Erwin P., Caon N., Trujillo I.,
      2001, ApJ, 563, L11
\bibitem[\protect\citeauthoryear{Gurvits}{1994}]{b11} Gurvits L.I., 1994, ApJ, 425, 442
\bibitem[\protect\citeauthoryear{Gurvits, Kellermann \& Frey}{1999}]{b12} Gurvits L.I.,
      Kellermann K.I., Frey S., 1999, A\&A, 342, 378
\bibitem[\protect\citeauthoryear{Guy et al.}{2005}]{b13} Guy J., Astier P., Nobili S.,
      Regnault N., Pain R., 2005, A\&A, 443, 781
\bibitem[\protect\citeauthoryear{Hamuy et al.}{1995}]{b48} Hamuy M., Phillips M.M., Maza J.,
      Suntzeff N.B., Schommer R.A., Avil\'es R., 1995, AJ, 109, 1
\bibitem[\protect\citeauthoryear{Hamuy et al.}{1996a}]{b45} Hamuy M., Phillips M.M.,
      Schommer R.A., Suntzeff N.B., Maza J., Avil\'es R., 1996a, AJ, 112, 2391
\bibitem[\protect\citeauthoryear{Hamuy et al.}{1996b}]{b46} Hamuy M., Phillips M.M.,
      Suntzeff N.B., Schommer R.A., Maza J., Avil\'es R., 1996b, AJ, 112, 2398
\bibitem[\protect\citeauthoryear{Henstock et al.}{1995}]{b14} Henstock D.R., Browne I.W.A.,
      Wilkinson P.N., Taylor G. B., Vermeulen R.C., Pearson T.J., Readhead A.C.S., 1995,
      ApJS, 100, 1 
\bibitem[\protect\citeauthoryear{Hoyle}{1959}]{b15} Hoyle F., 1959, in Bracewell R.,
      ed, Proc. IAU Symp. 9, Paris Symposium on Radio Astronomy.  Stanford Univ. Press,
      Stanford, p. 529
\bibitem[\protect\citeauthoryear{Jackson}{1973}]{b16} Jackson J.C., 1973, MNRAS, 162, 11{\small P}
\bibitem[\protect\citeauthoryear{Jackson}{2004}]{b17} Jackson J.C., 2004, JCAP, 11, 007
\bibitem[\protect\citeauthoryear{Jackson \& Dodgson}{1997}]{b18} Jackson J.C., Dodgson. M.,
      1997, MNRAS, 285, 806
\bibitem[\protect\citeauthoryear{Jackson \& Jannetta}{2006}]{b19} Jackson J.C., Jannetta A.L.,
      2006, JCAP, 11, 002 
\bibitem[\protect\citeauthoryear{Kellermann}{1972}]{b20} Kellermann K.I., 1972, AJ, 77, 531
\bibitem[\protect\citeauthoryear{Kellermann}{1993}]{b21} Kellermann K.I., 1993, Nat, 361, 134
\bibitem[\protect\citeauthoryear{Komatsu et al.}{2008}]{b22} Komatsu E. et al., 2008, ApJS, in press
      (arXive 0803.0547)
\bibitem[\protect\citeauthoryear{Legg}{1970}]{b23} Legg, T.H., 1970, Nat, 226, 65
\bibitem[\protect\citeauthoryear{Marscher \& Shaffer}{1980}]{b24} Marscher A.P., Shaffer D.B.,
      1980, AJ, 85, 668
\bibitem[\protect\citeauthoryear{Miley}{1971}]{b25} Miley G.K., 1971, MNRAS, 152, 477
\bibitem[\protect\citeauthoryear{Percival et al.}{2007}]{b26} Percival W. J., Cole S.,
      Eisenstein D.  J., Nichol R. C., Peacock J. A.,  Pope A. C., Szalay A. S., 2007,
      MNRAS, 381, 1053
\bibitem[\protect\citeauthoryear{Perlmutter et al.}{1999}]{b27} Perlmutter S. et al., 1999,
      ApJ, 517, 565
\bibitem[\protect\citeauthoryear{Pearson \& Readhead}{1981}]{b28} Pearson T.J., Readhead A.C.S.,
      1981, ApJ, 248, 61 
\bibitem[\protect\citeauthoryear{Pearson \& Readhead}{1988}]{b29} Pearson T.J., Readhead A.C.S.,
      1988, ApJ, 328, 114 
\bibitem[\protect\citeauthoryear{Pearson et al.}{1994}]{b30} Pearson T.J., Xu W., Thakkar D.D., 
      Readhead A.C.S., Polatidis A.G., Wilkinson P.N., 1994, in Kellermann K.I., Zenus J.A., eds,
      Proc. NRAO Workshop, Compact Extragalactic Radio Sources.
      National Radio Astronomy Observatory (NRAO), Green Bank, WV, p. 1
\bibitem[\protect\citeauthoryear{Phillips}{1993}]{b47} Phillips M.M., 1993, ApJ, 413, L105
\bibitem[\protect\citeauthoryear{Polatidis et al.}{1995}]{b31} Polatidis A.G., Wilkinson P.N.,
      Xu W., Readhead A.C.S., Pearson T.J., Taylor G.B., Vermeulen R.C., 1995, ApJS, 98, 1 
\bibitem[\protect\citeauthoryear{Preston et al.}{1985}]{b32} Preston R.A., Morabito D.D.,
      Williams J.G., Faulkner J., Jauncey D.L., Nicolson G.D., 1985, AJ, 90, 1599
\bibitem[\protect\citeauthoryear{Riess et al.}{1998}]{b33} Riess A.G. et al., 1998, AJ, 116, 1009
\bibitem[\protect\citeauthoryear{Riess et al.}{2004}]{b34} Riess, A.G. et al., 2004, ApJ, 607, 665
\bibitem[\protect\citeauthoryear{Riess et al.}{2007}]{b35} Riess, A.G. et al., 2007, ApJ, 659, 98
\bibitem[\protect\citeauthoryear{Schmidt et al.}{1998}]{b36} Schmidt B.P. et al., 1998, ApJ, 507, 46
\bibitem[\protect\citeauthoryear{Seo \& Eisenstein}{2003}]{b37} Seo H.-J., Eisenstein D.J., 2003,
      ApJ, 598, 720
\bibitem[\protect\citeauthoryear{Taylor et al.}{1994}]{b38} Taylor G.B., Vermeulen R.C.,
      Pearson T.J., Readhead A.C.S., Henstock D.R., Browne I.W.A., Wilkinson P.N., 1994,
      ApJS, 95, 345 
\bibitem[\protect\citeauthoryear{Taylor et al.}{1996}]{b39} Taylor G.B., Vermeulen R.C.,
      Readhead A.C.S., Pearson T.J., Henstock D.R., Wilkinson P.N., 1996, ApJS, 107, 37 
\bibitem[\protect\citeauthoryear{Thakkar et al.}{1995}]{b40} Thakkar D.D., Xu W., Readhead A.C.S.,
      Pearson T.J., Taylor G.B., Vermeulen R.C., Polatidis A.G., Wilkinson P.N., 1995,
      ApJS, 98, 33 
\bibitem[\protect\citeauthoryear{Wilkinson et al.}{1994}]{b41} Wilkinson P.N., Polatidis A.G.,
      Readhead A.C.S., Xu W., Pearson T.J., 1994, ApJ, 432, L87
\bibitem[\protect\citeauthoryear{Wood-Vasey et al.}{2007}]{b42} Wood-Vasey, W. M. et al., 2007,
      ApJ, 666, 694
\bibitem[\protect\citeauthoryear{Weinberg}{1972}]{b43} Weinberg S., 1972,
      Gravitation and Cosmology, Wiley, New York
\bibitem[\protect\citeauthoryear{Xu et al.}{1995}]{b44} Xu W., Readhead A.C.S., Pearson T.J.,
      Polatidis A.G., Wilkinson P.N., 1995, ApJS, 99, 297 
\end{thebibliography}
\end{document}